\documentclass[]{aa}        
\usepackage{graphicx}
\usepackage{bm}
\usepackage{txfonts}
\begin{document}
\def\bea{\begin{eqnarray}}
\def\eea{\end{eqnarray}}
\def\be{\begin{equation}}
\def\ee{\end{equation}}
\def\rra{\right\rangle}
\def\lla{\left\langle}
\def\le#1{\label{eq:#1}}
\def\re#1{\ref{eq:#1}}
\def\eps{\epsilon}
\def\sgm{\Sigma^-}
\def\la{\Lambda}
\def\kv{\bm{k}}
\def\vd{{\cal V}} 
\def\pp{{\cal P}} 
\def\kk{{\cal K}} 

\title{
The maximum and minimum mass of protoneutron stars\\
in the Brueckner theory}

\author{G. F. Burgio and H.-J. Schulze}

\institute{
INFN Sezione di Catania and 
Dipartimento di Fisica, Universit\'a di Catania,
Via Santa Sofia 64, I-95123 Catania, Italy}

\date{Received / Accepted}


\abstract{
We study the structure of protoneutron stars 
within the finite-temperature Brueckner-Bethe-Goldstone theoretical approach,
paying particular attention to how it is joined to a 
low-density nuclear equation of state.
We find a slight sensitivity of the minimum value of the protoneutron star
mass to the low-density equation of state,
whereas the maximum mass is hardly affected. 
}

\authorrunning{G. F. Burgio and H.-J. Schulze} 
\titlerunning{The maximum and minimum mass}


\maketitle

\section{Introduction}
\label{s:intro}

A lot of effort is currently dedicated to 
dynamical simulations of supernovae explosions 
(Burrows \& Lattimer 1986; Pons et al. 1999; Villain et al. 2004; 
Liebend\"orfer et al. 2005; Fischer et al. 2009)
and the subsequent formation and evolution of a protoneutron star (PNS).
The fundamental input to these calculations is the nuclear equation of state
(EOS) over a wide range of densities,
apart from microscopic information regarding diffusion and cooling processes.
The output are time-dependent radial profiles of the thermodynamic quantities 
of interest, such as temperature, entropy, particle fractions, etc.

According to current understanding,
one can very roughly identify different stages of this process
(Burrows \& Lattimer 1986; Pons et al. 1999; Strobel et al. 1999).
(i) For about 1 second after supernova core bounce, the system
consists of a relatively cool central region surrounded by a hot mantle, 
collapsing and radiating off neutrinos quickly,
while still also accreting material.
(ii) For the next 20 or so seconds one can identify a
slowly developing state of the proper PNS,
the system first deleptonizing and heating up the interior parts
of the star in the process,
and then beginning to cool down by further neutrino diffusion.
(iii) After several tens of seconds, the final state of a cold NS 
has basically been formed, 
which continues to cool down slowly first by neutrino and later 
by photon emission.

In this article we do not intend to perform dynamical simulations,
but focus on the consistent construction 
and evaluation of basic consequences
of the temperature-dependent nuclear EOS during the prominent PNS stage (ii).
We therefore assume strongly idealized, static profiles representing this
environment. For simplicity  we use 
a constant entropy per baryon throughout the star and
investigate the sensitivity of the results to the chosen value of entropy $S/A$,
as is usually done (Gondek et al. 1997; Goussard et al. 1997).

Another important aspect is the treatment of neutrino trapping.
Gravitational collapse calculations of the white-dwarf core of massive stars
indicate that, at the onset of trapping, the electron lepton fraction is
$Y_e = x_e + x_{\nu_e} \approx 0.4$,
the precise value depending on the efficiency of electron capture
reactions during the initial collapse stage.
Also, because no muons are present when neutrinos become trapped,
the constraint
$Y_\mu = x_\mu - x_{\bar{\nu}_\mu} = 0$
is imposed.
We fix the $Y_l$ at these values in our calculations 
for neutrino-trapped matter.
Furthermore, since low-density nuclear matter loses neutrino very fast,
the concept of a neutrino sphere,
inside which neutrinos are trapped, is often introduced during the PNS stage
(Gondek et al. 1997; Fischer et al. 2009). 
Our treatment of the neutrino sphere is explained later in section~\ref{s:low}.

One of the most advanced microscopic approaches to the EOS of 
nuclear matter is the Brueckner theory. 
In recent years, it has made rapid progress in several aspects.
(i) The convergence of the Brueckner-Bethe-Goldstone (BBG) expansion has been firmly
established (Song et al. 1998; Baldo et al. 2001). 
(ii) The addition of phenomenological three-body forces (TBF) based on 
the Urbana model (Carlson et al. 1983; Schiavilla et al. 1986),
permitted improving the agreement to a large extent with the empirical
saturation properties 
(Baldo et al. 1997; Zhou et al. 2004; Li et al. 2006,2008a).
(iii) The Brueckner-Hartree-Fock (BHF) approach has been extended in a fully microscopic and 
self-consistent way to describe nuclear matter also containing 
hyperons (Schulze et al. 1995, 1998). 
iv) It has also been extended to the finite-temperature regime  
within the Bloch-De Dominicis formalism with a realistic
nucleon-nucleon interaction (Baldo \& Ferreira 1999; Baldo 1999).

We used the finite-temperature BHF EOS to model PNSs
in our previous papers (Nicotra et al. 2006; Burgio \& Schulze 2009).
The present article represents further evolution in this approach, 
in particular improving two aspects of the previous work:
(a) an exact calculation of the finite-temperature EOS, 
in contrast to the so-called ``frozen correlations approximation''
used before; 
(b) particular attention to how to join the BHF high-density EOS
to different EOSs describing the low-density
domain of clustered nuclear matter.  

The structure of this article is as follows.
The BHF approach at finite temperature is reviewed in Sect.~2,
and the resulting composition and EOS of beta-stable matter are presented in
Sect.~3.
The joining to different low-density EOSs is discussed in Sect.~4,
and the final properties of PNSs are shown in Sect.~5.
Conclusions are drawn in Sect.~\ref{s:sum}.

\section{Brueckner theory at finite temperature}
\label{s:bd}

The formalism that is closest to the BBG expansion, 
and that actually reduces to it in the zero-temperature limit, 
is the one formulated by Bloch \& De Dominicis (1958, 1959a, 1959b).
In this approach, the essential ingredient is the two-body scattering 
matrix $K$, which, along with the single-particle potential $U$, 
satisfies the self-consistent equations
\bea
  \lla k_1 k_2 | K(W) | k_3 k_4 \rra
 &=& \lla k_1 k_2 | V | k_3 k_4 \rra  
\nonumber\\&&
 \hskip -27mm +\; \mathrm{Re} \sum_{k_3' k_4'} 
 \langle k_1 k_2 | V | k_3' k_4' \rangle 
 { [1-n(k_3')] [1-n(k_4')] \over 
   W - E_{k_3'} - E_{k_4'} + i\eps }
 \langle k_3' k_4' | K(W) | k_3 k_4 \rangle
\nonumber\\&&
\label{eq:kkk}
\eea
and
\be
 U(k_1) = \sum_{k_2} n(k_2) \lla k_1 k_2 | K(W) | k_1 k_2 \rra_A \:,
\label{eq:ueq}
\ee
where $k_i$ generally denote momentum, spin, and isospin.
Here $V$ is the two-body interaction, 
$W = E_{k_1} + E_{k_2}$ represents the starting energy, and
$E_k = k^2/2m + U(k)$ the single-particle energy.
Equation~(\ref{eq:kkk}) coincides with the Brueckner equation for the 
$K$ matrix at zero temperature 
if the single-particle occupation numbers $n(k)$ are taken at $T = 0$. 
At finite temperature, $n(k)$ is a Fermi distribution.
For a given density and temperature, 
Eqs.~(\ref{eq:kkk}) and (\ref{eq:ueq}) have to be solved 
self-consistently along with the following equation for the auxiliary
chemical potential $\tilde{\mu}$,
\be
 \rho = \sum_k n(k) = 
 \sum_k {1\over e^{\beta (E_k - \tilde{\mu})} + 1 } \:.
\label{e:ro}
\ee

The grand-canonical potential density $\omega$ in the Bloch-De Dominicis 
framework can be written as the sum of a mean-field term
and a correlation contribution
(Baldo \& Ferreira 1999; Baldo 1999), 
\bea
 \omega &=& -\sum_k \left[ {1\over\beta} 
 \ln\left( 1 +  e^{-\beta (E_k - \tilde{\mu})}\right) 
 + n(k) U(k) \right]
\nonumber
\\&&+
 {\frac{1}{2}} \int {dW\over 2\pi} e^{\beta(2\tilde{\mu}-W)} 
 \,{\rm Tr_2}\bigg(\!\arctan\big[{\cal{K}}(W) \pi\delta(H_0-W) \big] \bigg) \:, 
\label{e:om}
\eea
where the trace, $\rm Tr_2$, is taken in the space of
antisymmetrized two-body states, and the two-body scattering matrix $\cal K$
is defined by
\bea
  \lla k_1 k_2 | {\cal{K}}(W) | k_3 k_4 \rra  &=& 
  \lla k_1 k_2 | K(W) | k_3 k_4 \rra  
  \prod_{i=1,4}\! \sqrt{1-n_i(k)} \; .
\eea

If one only considers the first term in the expansion of the arctan function
(Baldo \& Ferreira 1999, see Appendix A), 
then the correlation term reduces to
\be
 \omega_c =  
 {\frac{1}{2}} \sum_{k} n(k) U(k) \:,
\label{e:om1}
\ee
which defines the grand-canonical potential in total analogy with the BBG 
binding potential,
just using Fermi functions instead of the usual step functions 
at zero temperature.
In this way one neglects a series of terms proportional to
\hbox{$n(k)[1-n(k)]$} (or powers of it), 
which turn out to be negligible in the
temperature and density ranges relevant for neutron and protoneutron stars.

A further simplification can be achieved by disregarding the effects of 
finite temperature on the single-particle potential $U$, 
and using the $T=0$ results 
in order to speed up the calculations 
(frozen correlations approximation).
This was the procedure followed in our previous publications
(Nicotra et al. 2006; Burgio \& Schulze 2009). 
On the contrary,
all calculations shown in this paper are obtained without 
introducing any approximation on the single-particle potential $U$,
and using the exact expression for the grand-canonical potential, 
Eq.~(\ref{e:om}). 

Furthermore, in the previous work the 
calculations were performed using the $G$-matrix instead of the $K$-matrix, 
[i.e., taking the real part in Eq.~(\ref{eq:ueq}) instead of
Eq.~(\ref{eq:kkk})], 
which produces a more repulsive EOS at high density.
Strictly speaking, the Bloch-De Dominicis formalism has been derived
for the $K$-matrix; on the other hand, the use of the $G$-matrix
ensures the compatibility with the usually performed continuous-choice 
BHF calculations at zero temperature.

In this framework, the free energy density is then 
\be
 f = \omega + \rho \tilde{\mu} \:,
\label{e:f}
\ee
and all other thermodynamic quantities of interest can be computed from it; 
namely, the ``true" chemical potential $\mu$, pressure $p$, 
entropy density $s$, and internal energy density $\eps$ read as
\bea
 \mu &=& {{\partial f}\over{\partial \rho}} \:,
\\
 p &=& \rho^2 {{\partial (f/\rho)}\over{\partial \rho} }  
 = \mu \rho - f \:,
\\
 s &=& -{{\partial f}\over{\partial T}} \:,
\\
 \eps &=& f + Ts \:.
\eea  
We stress that this procedure permits to fulfill the Hugenholtz-Van Hove 
theorem in the calculation of thermodynamical quantities in the Brueckner theory.
For an extensive discussion of this topic, the reader is referred 
to Baldo (1999), and references therein.

It is well known that at zero temperature the non-relativistic 
microscopic approaches do not correctly reproduce the nuclear matter 
saturation point
$\rho_0 \approx 0.17~\mathrm{fm}^{-3}$, $E/A \approx -16$ MeV. 
One common way of correcting this deficiency 
is to introduce three-body forces among nucleons. 
Since a complete microscopic theory of
TBF is not available yet, we have adopted the phenomenological Urbana model 
(Carlson et al. 1983; Schiavilla et al. 1986),
which consists of an attractive term due to two-pion exchange with 
excitation of an intermediate $\Delta$ resonance 
and a repulsive phenomenological central term.
In the BBG approach, the TBF is reduced to a density-dependent two-body force 
by averaging over the position of the third particle, 
assuming that the probability of having two particles at a
given distance is reduced according to the two-body correlation function.
The corresponding EOS reproduces the nuclear matter saturation point correctly 
(Baldo et al. 1997; Zhou et al. 2004; Li et al. 2008a) 
and fulfills several requirements from the nuclear phenomenology.
In all calculations presented in this paper we used the Argonne $V_{18}$ 
nucleon-nucleon potential (Wiringa et al. 1995) 
along with the phenomenological Urbana TBF.

\begin{figure}[t]
\centering
\includegraphics[height=120mm,angle=0,clip]{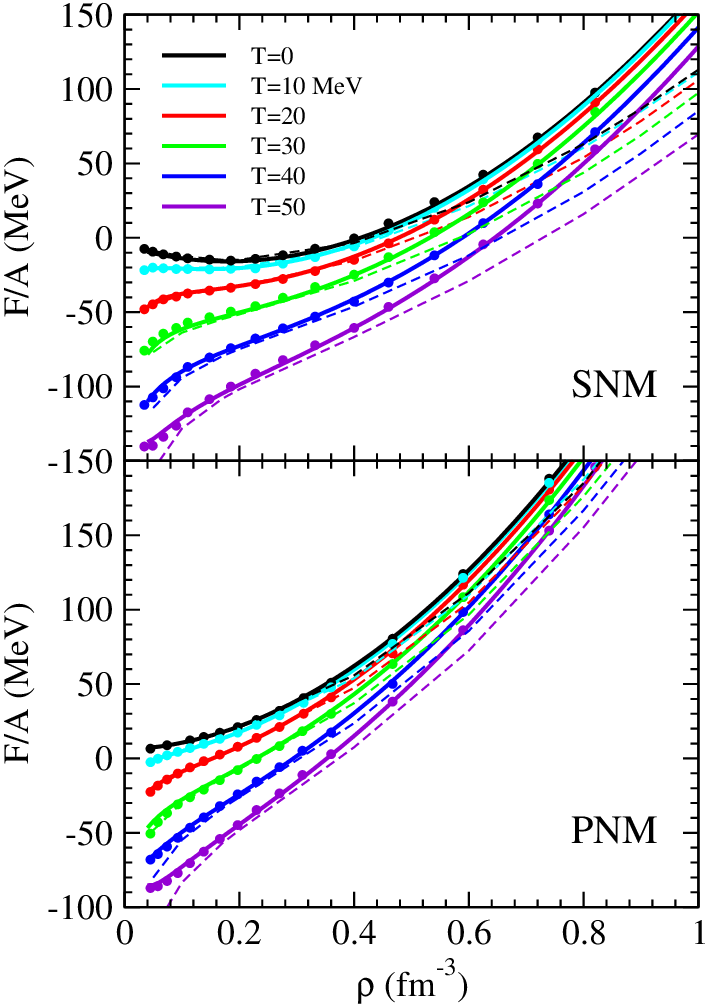}
\caption{
Free energy per nucleon for symmetric (upper panel) and 
pure neutron (lower panel) matter (solid lines).
The temperatures vary from 0 to 50 MeV in steps of 10 MeV.
See text for details.}
\label{f:ba}
\end{figure}

\begin{figure}[t]
\centering
\includegraphics[height=120mm,angle=0,clip]{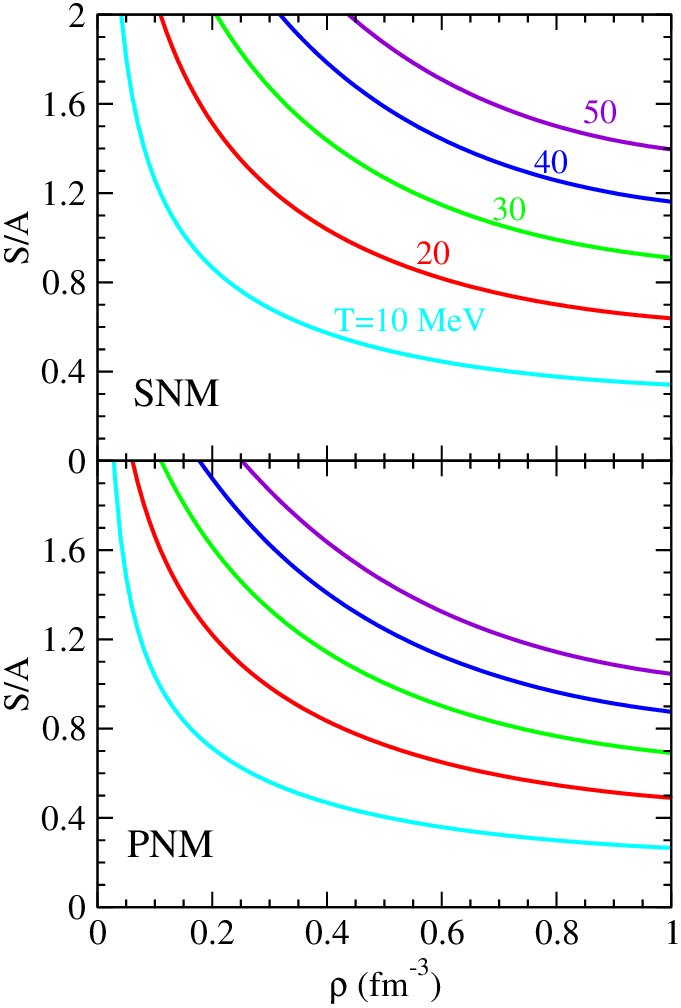}
\caption{
The entropy per particle for symmetric (upper panel) and 
pure neutron (lower panel) matter 
at different temperatures.}
\label{f:sa}
\end{figure} 

In Fig.~\ref{f:ba} we display the free energy vs.~the nucleon density,
obtained following the procedure discussed above, 
both for symmetric and pure neutron matter, 
and for several values of the temperature between 0 and 50 MeV. 
The dots represent the numerical results,
and the curves are empirical fits, 
which can be easily implemented in numerical simulations. 
The thin dashed curves represent the calculations obtained with the 
$G$-matrix and the frozen correlations approximation for the
single-particle potential $U$, as adopted in our previous papers
(Nicotra et al. 2006; Burgio \& Schulze 2009). 
The discrepancy at high density is mainly due to the use of the $G$-matrix 
instead of the $K$-matrix, whereas the
differences observed at low density and high temperature 
are consequences of the frozen correlations approximation.
  
We find that the following functional forms provide excellent parametrizations
of the new numerical results in the required ranges of density 
($0.03\;{\rm fm}^{-3} \lesssim \rho \lesssim 1\;{\rm fm}^{-3}$)
and temperature ($0\;{\rm MeV} \leq T \leq 50\;{\rm MeV}$)
for symmetric (SNM) and pure neutron matter (PNM):
\bea
 {F\over A}(T,\rho) &=& 
 -(137+157 t^2)\rho + 308 \rho^{1.82} + 207 t^2 \ln(\rho) 
\nonumber\\
 && + (-47.5 t^2 + 71 t^{2.41})/\rho - 5 \:, 
\quad\quad \rm(SNM)\:, 
\label{e:fitfs}
\\
 {F\over A}(T,\rho) &=& 
 (11 - 122 t^2)\rho + 309 \rho^{1.95} + 173 t^2 \ln(\rho) 
\nonumber\\
 && + (-48 t^2 + 71 t^{2.35})/\rho + 6\:, 
\quad\quad \rm(PNM) \:,
\label{e:fitfn}
\eea
where $t=T/(100\;\mathrm{MeV})$ and $F$ and $\rho$ are given in
MeV and fm$^{-3}$, respectively.

We notice that the free energy of symmetric matter shows a typical 
Van der Waals behavior (with $T_C\approx 19$ MeV, 
$\rho_c \approx 0.06~\rm fm^{-3}$)
and is a monotonically decreasing function of the temperature.  
At $T=0$ the free energy coincides with the total energy and the corresponding 
curve is just the usual nuclear matter saturation curve.

From the free energy we can calculate the entropy  
(in units of the Boltzmann constant) 
from the thermodynamical relationship
\be
 \frac{S}{A}(\rho) = -\Big( \frac{\partial F/A}{\partial T} \Big)_\rho \:,
\ee
and this is displayed in Fig.~\ref{f:sa} for both SNM and PNM. 

It turns out that the dependence of the free energy on the proton fraction 
can be approximated very well by a quadratic dependence, as
at zero temperature (Bombaci \& Lombardo 1991; Baldo et al. 1998, 2000a):
\be
 {F\over A}(T,\rho,x) \approx 
 {F\over A}(T,\rho,x=0.5) + (1-2x)^2 F_\mathrm{sym}(T,\rho) \:,
\label{e:parab}
\ee
where the symmetry energy $F_\mathrm{sym}$ can be expressed in
terms of the difference of the energy per particle between pure neutron 
($x=0$) and symmetric ($x=0.5$) matter:
\bea
  F_\mathrm{sym}(T,\rho) &=& 
  - {1\over 4} {\partial(F/A) \over \partial x}(T,\rho,0)
\\
  &\approx& {F\over A}(T,\rho,0) - {F\over A}(T,\rho,0.5) \:.
\label{e:sym}
\eea
Therefore, in order to treat the beta-equilibrated case, 
it is only necessary to use the parametrizations 
Eqs.~(\ref{e:fitfs},\ref{e:fitfn})
of the free energy for SNM and PNM.

\begin{figure}[t]
\centering
\includegraphics[height=65mm,angle=0,clip]{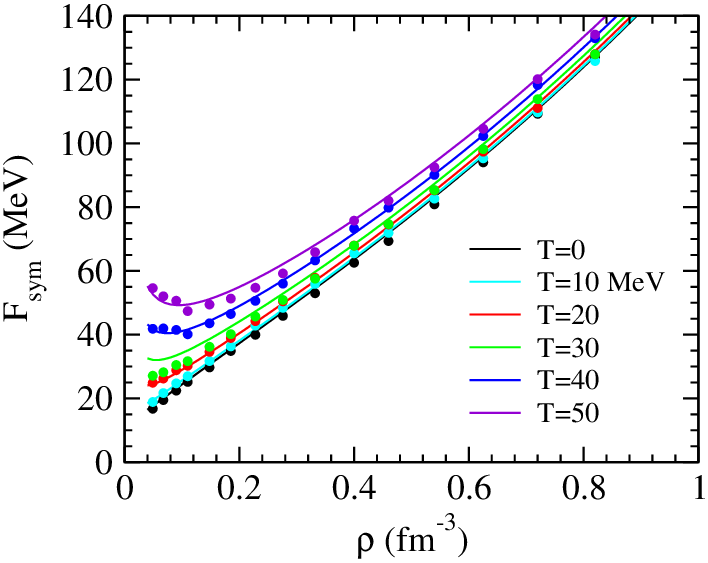}
\caption{
The free symmetry energy as a function of the nucleon density
at different temperatures.}
\label{f:fsym}
\end{figure} 

For completeness, we display in Fig.~\ref{f:fsym} the behavior of
the free symmetry energy for several values of the temperature. 
Full dots represent the numerical values, 
and the solid curves are the results of the polynomial fits.
Even at finite temperature, the symmetry energy increases nearly 
linearly with density, as is generally the case 
in the BHF approach (Li et al. 2006,2008a).
However, at low density there are significant deviations due to finite
temperature, which might have experimental implications.
The symmetry energy plays a significant role not only in the cooling
of protoneutron stars, but also in the dynamics of heavy ion collisions 
induced by radioactive beams and the structure of exotic nuclei.
For a review, see Li et al. (2008b).

\section{Composition and EOS of hot stellar matter}

\begin{figure*}[t] 
\centering
\includegraphics[width=170mm,clip]{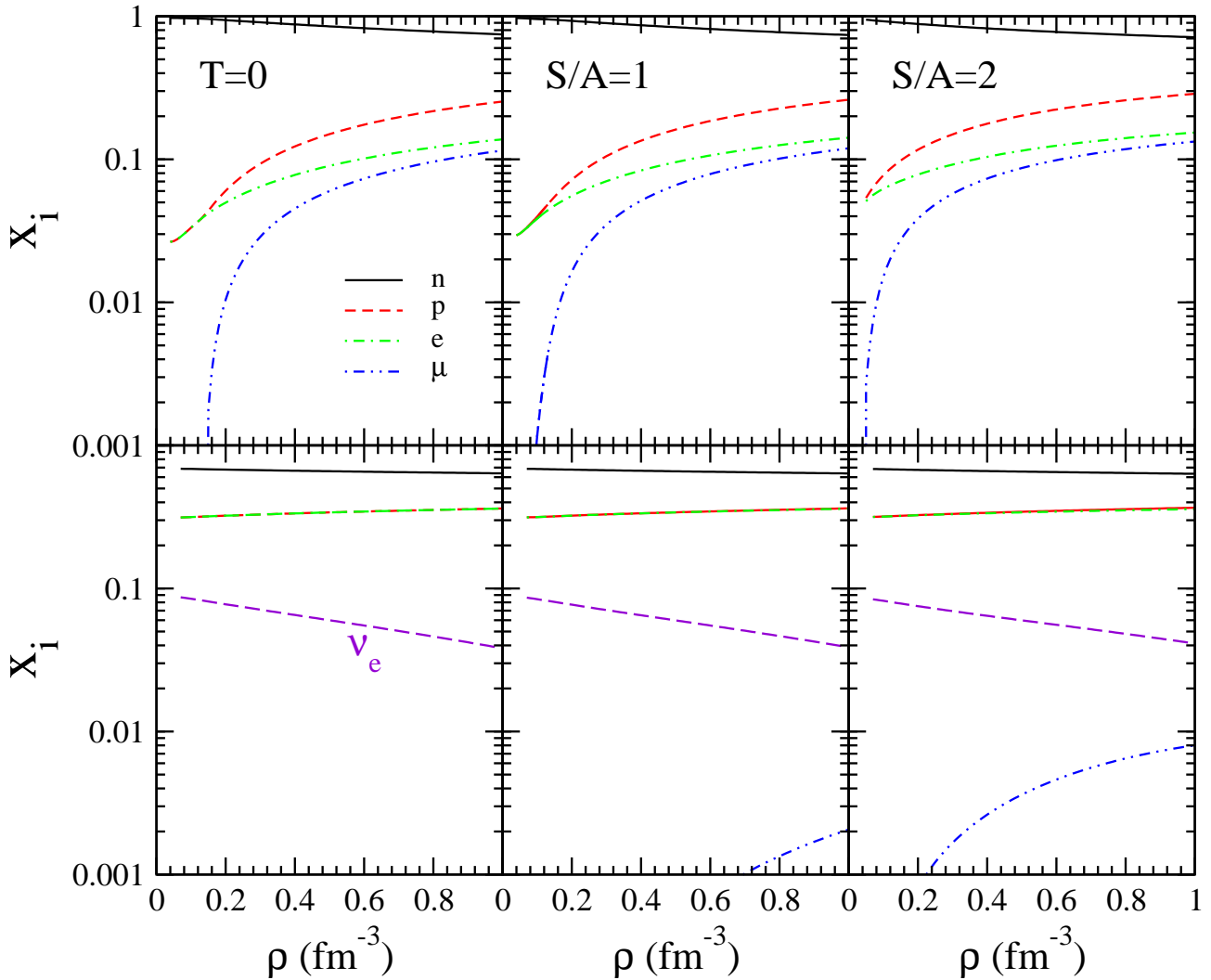}
\caption{
Relative populations as functions of baryon density
in neutrino-free (upper panels), and neutrino-trapped (lower panels) 
beta-equilibrated matter at entropies $S/A = 0, 1, 2$.}
\label{f:xi}
\end{figure*} 

In neutrino-trapped $\beta$-stable nuclear matter,
the chemical potential of any particle $i=n,p,l$ is uniquely determined
by the conserved quantities baryon number $B_i$, electric charge $Q_i$,
and weak charges (lepton numbers) $L^{(e)}_i$, $L^{(\mu)}_i$:
\be
 \mu_i = B_i\mu_n - Q_i(\mu_n-\mu_p)
 + L^{(e)}_i\mu_{\nu_e}  + L^{(\mu)}_i\mu_{\nu_\mu} \:.
\label{e:mufre}
\ee
For stellar matter containing nucleons and leptons as relevant degrees
of freedom, the chemical equilibrium conditions read explicitly as
\be
 \mu_n - \mu_p = \mu_e - \mu_{\nu_e} = \mu_\mu + \mu_{\bar{\nu}_\mu} \:.
\label{e:beta}
\ee
At given baryon density $\rho$,
these equations have to be solved with the
charge neutrality condition
\be
 \sum_i Q_i x_i = 0
\label{e:neutral}
\ee
and those expressing conservation of lepton numbers
\be
 Y_l = x_l - x_{\bar l} + x_{\nu_l} - x_{\bar{\nu}_l}
 \:,\quad l=e,\mu \:.
\label{e:lepfrac}
\ee
As discussed in Sect.~\ref{s:intro}, 
we fix the lepton fractions to $Y_e=0.4$ and $Y_\mu=0$ 
for neutrino-trapped matter.
When the neutrinos have left the system,
their partial densities and chemical potentials vanish
and the above equations simplify accordingly.

The nucleon chemical potentials are obtained from the
free energy density $f$, Eq.~(\ref{e:f}),
\bea
 \mu_i(\{\rho_j\}) &=&
 \left. \frac{\partial f}{\partial \rho_i} \right|_{\rho_{j\neq i}} \:,
 \ i=n,p \:,
\label{mun:eps}
\eea
and the chemical potentials of the non-interacting leptons are obtained 
by solving numerically the free Fermi gas model at finite temperature. 
Once the hadronic and leptonic chemical potentials are known,
one can proceed to calculate the composition of the 
$\beta$-stable stellar matter,
and then the total pressure $p$ through the usual thermodynamical relation
\be
 p = \rho^2 {\partial{(f/\rho)}\over \partial{\rho}}
 = \sum_i \mu_i \rho_i - f  \:.
\ee

Following this procedure,
we first discuss the populations of beta-equilibrated stellar matter.
In Fig.~\ref{f:xi} we display the relative particle fractions as a function 
of the baryon density for cold (left panels) and hot beta-stable matter 
characterized by entropy values 
$S/A=1$ (central panels) and $S/A=2$ (right panels). 
The upper panels show the particle fractions when stellar matter contains 
only neutrons, protons, electrons, and muons, whereas 
the lower panels show the particle fractions in neutrino-trapped matter. 
We notice that the electron fraction is greater in neutrino-trapped 
matter than in the neutrino-free case; 
therefore, the nuclear matter is more symmetric, 
and the resulting EOS will be softer.
The appearance of muons in trapped matter is shifted to higher density values, 
because their onset is determined by the difference between the 
neutron and proton chemical potentials, Eq.~(\ref{e:beta}).

We observe that thermal effects influence
the populations mainly in the low-density region. 
In fact, the presence of tails in the Fermi distribution makes it possible 
to create (anti)particles at any density and thus typical production thresholds,
like for muon creation, disappear at finite entropy.
At high density, thermal effects are less important and
do not change the composition appreciably when increasing entropy. 
As already found in (Nicotra et al. 2006) in the isothermal case 
and in (Burgio \& Schulze 2009),
the proton fraction increases slightly at high density with increasing entropy,
thus causing a softening of the EOS. 
This is at variance with the results 
shown in (Prakash et al. 1997), and is presumably caused by the different 
temperature dependence of the potential part of the EOS.
 
\begin{figure}[t] 
\centering
\includegraphics[height=10cm,clip]{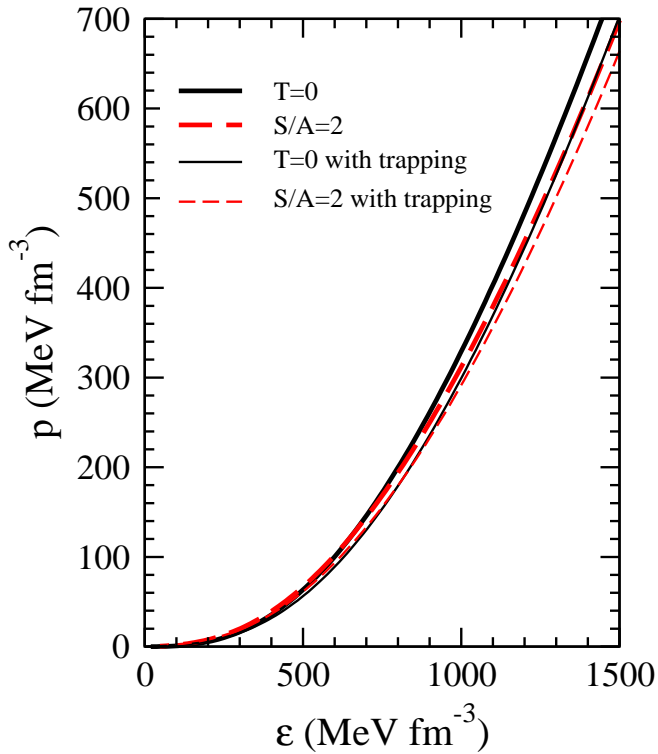}
\caption{
Pressure as a function of energy density for beta-equilibrated 
cold and hot matter with and without neutrino trapping.}
\label{f:eos}
\end{figure} 

In Fig.~\ref{f:eos}, we display the pressure for beta-stable asymmetric matter
with and without neutrinos, 
as a function of the energy density at entropy $S/A=2$ (red curves),
in comparison with cold beta-stable matter (black curves).
As discussed above, we notice that both thermal effects and neutrino trapping
produce a softer EOS than in the cold case.

\section{Low-density EOS}
\label{s:low}

The Brueckner approach provides a realistic modeling of nuclear matter
only at densities above about half normal nuclear matter density.
Below this threshold, clusterization sets in and the system becomes
inhomogeneous.
Another theoretical approach therefore has to be used in this ``low-density''
regime, and we employ two widely used liquid-drop-type
models at finite temperature, 
namely the one of Lattimer \& Swesty (1997)
and the one of Shen et al. (1998a, 1998b).
In the first case we in fact compare
a model with a (too) large compressibility, $K=370$ MeV, here denoted by ``LS'' 
and a second one with a lower compressibility, $K=263$ MeV, denoted by ``SKa''.
The Shen EOS is characterized by a compressibility $K=281\;\rm MeV$ 
and a symmetry energy at saturation of $E_{\rm sym}=36.9\;\rm MeV$,
while the BHF EOS has $K = 210\;\rm MeV$ and
$E_{\rm sym}=34.7\;\rm MeV$.

Of course, since no phase transition is involved,
but only two different theoretical descriptions of the same state of matter,
the joining of the two EOSs requires the thermodynamical observables, 
i.e., free energy, internal energy, pressure, and chemical potentials,
to be continuous functions of the baryon density.
Of particular interest are therefore 
(i) the consistent joining of the low- and high-density EOS, and 
(ii) eventual differences caused by the use of different low-density models.
These aspects are studied in detail in the following. 

\begin{figure}[t]
\centering
\includegraphics[width=85mm,clip]{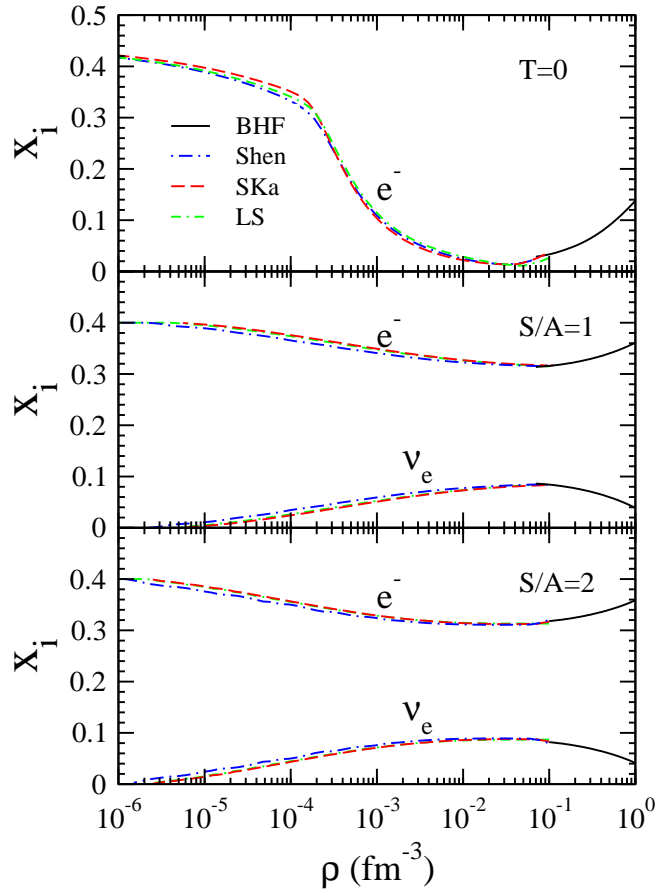}
\caption{
The lepton fractions in beta-stable matter
as function of baryon density for the different EOSs.}
\label{f:xl}
\end{figure}

Another important feature of the low-density domain
is the treatment of neutrino trapping.
Physically, neutrinos escape rapidly from the low-density matter
during the PNS evolution,
and so the trapping condition $Y_e=0.4$ does not apply anymore.
This effect can be roughly modeled by the concept of a neutrino sphere
inside which the neutrinos are trapped.
Typical model-dependent values for the location of the neutrino sphere
found in the literature are
$2\times10^{-3}\; \rm fm^{-3}$ (Gondek et al. 1997),
$6\times10^{-4}\; \rm fm^{-3}$ (Strobel et al. 1999),
and $2\times10^{-5}\; \rm fm^{-3}$ (Fischer et al. 2009).
Given these variations, 
we choose the following ``natural cutoff'' procedure: By
imposing the condition $Y_e=0.4$ at any density, at a certain 
threshold density $\rho_\nu \approx 10^{-5}-10^{-6}\; \rm fm^{-3}$,  
the electron fraction $x_e$ becomes 0.4, 
and the neutrinos disappear naturally.
For lower densities we consider the matter untrapped.
This is illustrated in Fig.~\ref{f:xl}, showing the electron and
neutrino fractions following this recipe with the different EOSs.
This simple procedure avoids making assumptions about the neutrino sphere,
but clearly a satisfactory treatment of neutrino trapping
would require coupled dynamical calculations of PNS evolution
and the temperature-dependent EOS,
which might be possible in the future.

\begin{figure}[t]
\centering
\includegraphics[width=85mm,clip]{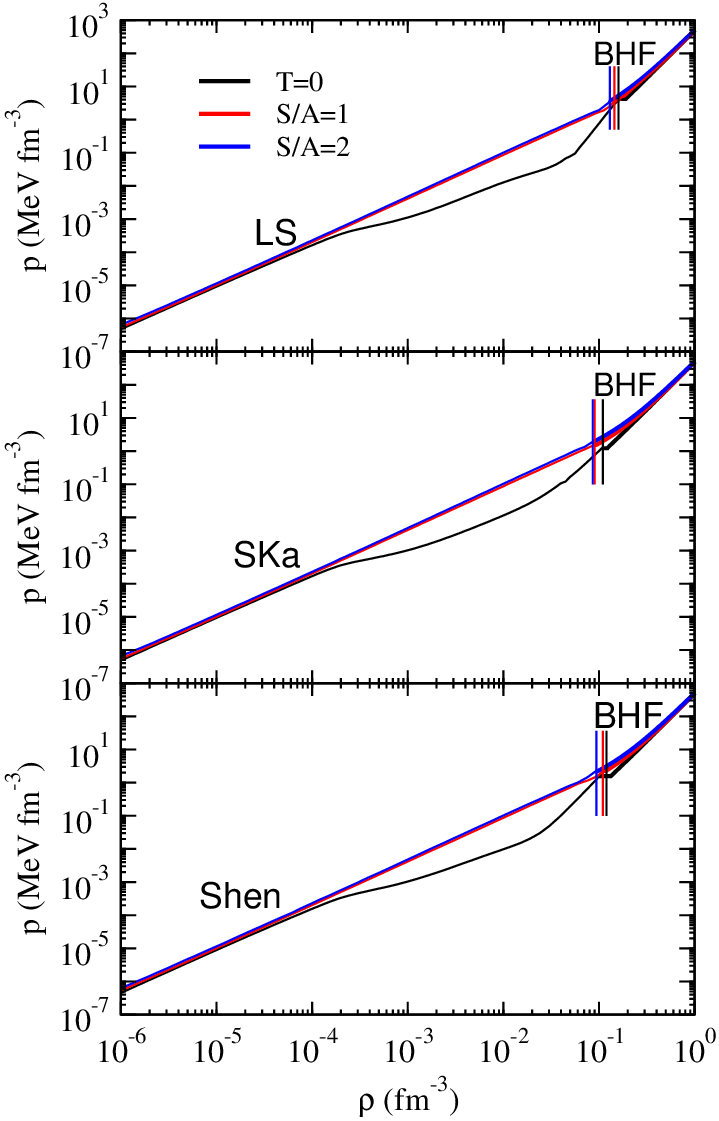}
\caption{
The EOS for cold untrapped matter (black curves) and for neutrino-trapped
matter with entropy $S/A=1$ (red curves) and $S/A=2$ (blue curves).}
\label{f:eos_ls}
\end{figure}

In Fig.~\ref{f:eos_ls}, we display 
the joint low-density LS, SKa, Shen + BHF EOSs over a wide range of densities.
The black lines represent the calculations performed at $T=0$, 
whereas the red (blue) lines are the ones for fixed entropy $S/A=1\ (S/A=2)$
in neutrino-trapped matter. 
The transition densities between low-density inhomogeneous regime
and homogeneous BHF regime in the different configurations 
are displayed in Fig.~\ref{f:eos_ls} as vertical bars,
and summarized in Table~\ref{t:trans}.
One observes in general a reduction of the transition density
with decreasing compressibility of the low-density EOS
and with increasing temperature.
In all configurations, a consistent joining of the two
segments of the EOS is possible, fulfilling the above-mentioned 
continuity requirements. 
In practice we performed a Maxwell construction by equating pressure and 
chemical potentials of the low- and high-density sectors,
and verified that the other thermodynamic variables do not exhibit 
significant discontinuities at the transition point.

As a further general prominent feature, 
we notice a strong stiffening of the EOS in neutrino-trapped
matter as compared to the cold untrapped case 
in the domain $\rho \approx 10^{-4},\ldots,10^{-1}\; \rm fm^{-3}$ 
(black vs.~colored curves in Fig.~\ref{f:eos_ls}),
which comes from the greater lepton Fermi pressure in trapped matter
(larger electron fraction and additional neutrino contribution,
cf., Fig.~\ref{f:xl}).

\begin{table}[b]
\caption{Transition densities (in fm$^{-3}$)
between low-density (LS, SKa, Shen) and BHF EOS.}
\begin{tabular}{l|lll}
   & untrapped, T=0 & trapped, S/A=1 & trapped, S/A=2 \\
\hline
 LS   & 0.160  & 0.145  & 0.130 \\
 SKa  & 0.110  & 0.090  & 0.086 \\
 Shen & 0.120  & 0.110  & 0.094 \\
\end{tabular}
\label{t:trans}
\end{table}

\section{(Proto)neutron star configurations}

The stable configurations of a (P)NS can be obtained from the
well-known hydrostatic equilibrium equations
of Tolman, Oppenheimer, and Volkov (Shapiro \& Teukolsky 1983)
for pressure $p(r)$ and enclosed mass $m(r)$,
\bea
 {dp\over dr} &=& -\frac{Gm\eps}{r^2}
 \frac{ \big( 1 + p/\eps \big) \big( 1 + 4\pi r^3p/m \big)}
 {1-2Gm/r} \:,
\label{tov1:eps}
\\
 \frac{dm}{dr} &=& 4\pi r^{2}\eps \:,
\label{tov2:eps}
\eea
once the EOS $p(\eps)$ is specified, with
$\eps=\eps_N+\eps_L$
the total internal energy density ($G$ is the gravitational constant).
For a given central value of the energy density, the numerical integration of
Eqs.~(\ref{tov1:eps}) and (\ref{tov2:eps}) provides the mass-radius relation.

\begin{figure}[t]
\centering
\includegraphics[width=85mm,clip]{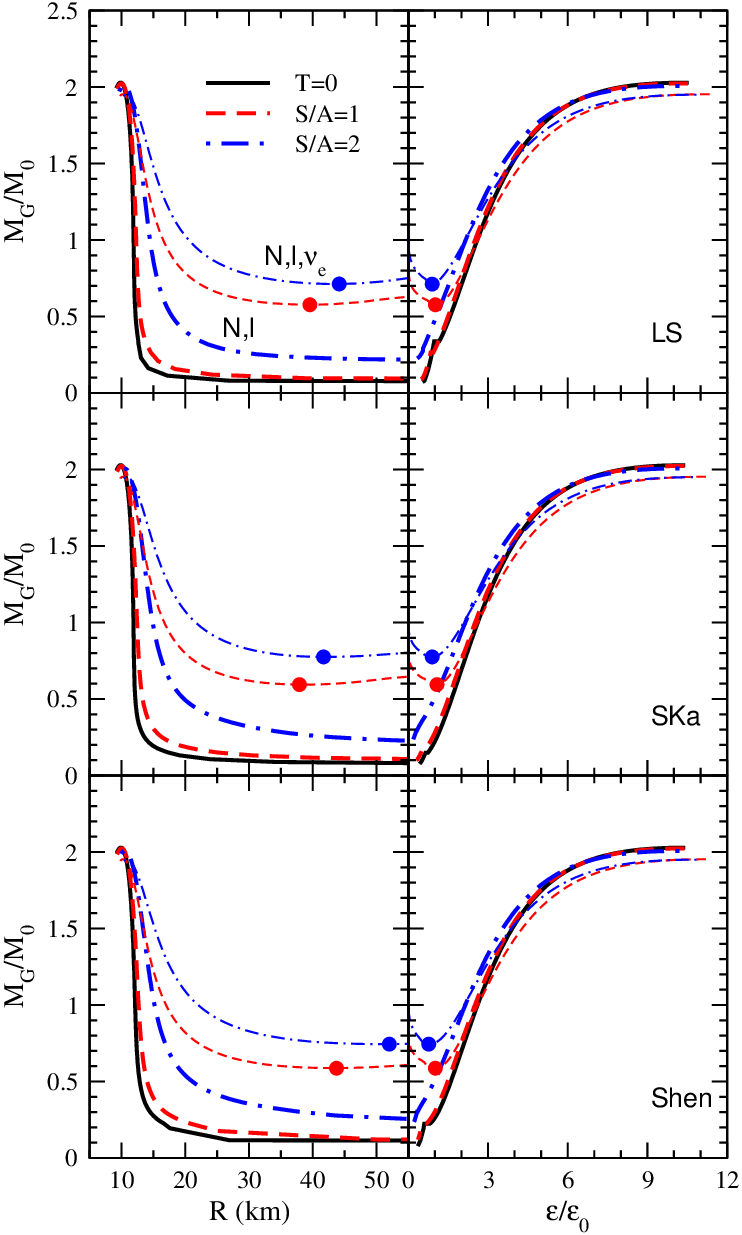}
\caption{
The gravitational mass 
as a function of the radius (left panels), 
and the central energy density (right panels),
obtained with the LS (top panels), SKa (central panels), and 
Shen (bottom panels) EOSs.}
\label{f:mr_ls}
\end{figure}

Our results, using the different EOSs introduced in the previous section,
are displayed in Fig.~\ref{f:mr_ls},
which shows the gravitational mass 
(in units of the solar mass $M_\odot=1.98\times 10^{33}$g) 
as a function of the radius (left panels), 
and the central density (right panels) 
(normalized with respect to the saturation value 
$\epsilon_0=152\;\rm MeV~fm^{-3}$). 
The black curves represent the calculations performed at $T=0$, 
whereas the thick (thin) colored lines denote stable configurations 
in neutrino-free (neutrino-trapped) matter
at constant entropy $S/A=1$ (dashed lines) and $S/A=2$ (dot-dashed lines).

We notice that the value of the maximum mass decreases slightly 
in neutrino-trapped matter, because of the softening of the EOS 
discussed before.
The value of the maximum mass turns out to be practically independent 
of the entropy, 
whereas the value of the radius for a fixed gravitational mass
depends strongly on the entropy and on the presence of neutrinos.  
The dots mark the values of the minimum mass and the corresponding radius
a PNS may have, 
with typical values in the range 0.6--0.77 $M_\odot$ and about 40--50 km, 
according to the chosen low-density EOS. 
The proper values are summarized in Table~\ref{t:mass}.
One notes that the minimum mass configurations only probe the low-density
part of the EOS. 
As expected, the minimum mass increases 
with decreasing compressibility of the EOS,
and with increasing entropy of the matter.

Our values of the minimum mass are very similar to the ones 
found in Gondek et al. (1997), 
where the LS EOS was used over the full density range of the PNS.
On the other hand, our values turn out to be
lower than those calculated in Strobel et al. (1999), 
the reason being
that our minimum mass is calculated during stage (ii) in the PNS evolution,
when neutrinos are still trapped, and the core and the mantle are 
characterized by the same entropy values as described in the Introduction.

\begin{table}
\setlength{\tabcolsep}{3pt}
\caption{Properties of (P)NS minimum and maximum mass configurations.}
\begin{tabular}{l|l|ccc|ccc}
  \multicolumn{2}{l|}{} & \multicolumn{3}{c|}{minimum mass} 
                        & \multicolumn{3}{c}{maximum mass} \\
  \multicolumn{2}{l|}{} & $M/M_\odot$ & $R$ (km) & $\rho_c/\rho_0$
                        & $M/M_\odot$ & $R$ (km) & $\rho_c/\rho_0$ \\
\hline
 untrapped  & LS   &  & &             & 2.03 & 9.86 & 10.55 \\
 T=0        & SKa  &  & &             & 2.03 & 9.86 & 10.42 \\
            & Shen &  & &             & 2.03 & 9.93 & 10.42 \\
\hline
 trapped    & LS   & 0.58 & 40 & 1.02 & 1.95 & 10.2 & 11.34 \\
 S/A=1      & SKa  & 0.60 & 38 & 1.08 & 1.95 & 10.2 & 11.20 \\
            & Shen & 0.58 & 44 & 1.02 & 1.95 & 10.3 & 11.20 \\
\hline
 trapped    & LS   & 0.70 & 44 & 0.90 & 1.95 & 10.7 & 10.85 \\
 S/A=2      & SKa  & 0.77 & 42 & 0.90 & 1.95 & 10.8 & 10.70 \\
            & Shen & 0.75 & 52 & 0.77 & 1.95 & 10.8 & 10.80 \\
\end{tabular}
\label{t:mass}
\end{table}

\section{Conclusions}
\label{s:sum}

We have presented a microscopic calculation of the EOS of hot nuclear
matter in the BHF approach
and provided convenient parametrizations of the free energy as a function
of baryon density, proton fraction, and temperature. All 
other thermodynamic quantities of interest can be derived from these.
This EOS was joined to different standard EOSs describing the low-density,
inhomogeneous domain of hot nuclear matter,
and we roughly modeled the structure of PNSs,
using strongly idealized temperature and neutrino-trapping profiles.

Altogether we find only small variations in our results for the minimum 
mass configurations when different low-density EOSs are employed, 
while maximum masses are practically unaffected.
Both neutrino trapping and (to a minor degree) finite temperature 
decrease the asymmetry of nuclear matter, 
and thus soften the EOS and decrease the maximum mass.

However, to provide more quantitative results, it will be 
necessary to perform consistent dynamical simulations of the PNS evolution, 
using self-consistent temperature and neutrino-trapping profiles, 
along with the temperature dependent EOS.
This is a difficult task for the future.

\section{Acknowledgements}

We thank M. Baldo, 
V. Ferrari, and L. Gualtieri 
for useful discussions. 
We acknowledge the support of COMPSTAR, 
a research and training program of the European Science Foundation.


\end{document}